\documentclass[journal=jctcce,manuscript=article,layout=twocolumn]{achemso}

\usepackage[version=3]{mhchem} 

\usepackage{graphicx} 
\usepackage{xcolor}
\usepackage{subcaption}
\usepackage{amsmath}  
\usepackage{amsthm}   
\usepackage{amssymb}
\usepackage{geometry}
\usepackage{bbold}
\usepackage{macros}
\usepackage{tabularx}

\mciteErrorOnUnknownfalse

\theoremstyle{plain}
\newtheorem{theorem}{Theorem}

\theoremstyle{remark}

\author{Yehor Tuchkov}
\affiliation{Department of Physics, Johannes Gutenberg University Mainz, Germany}
\alsoaffiliation{Flatiron Institute, Center for Computational Biology, New York, USA}

\author{Luke Evans}
\affiliation{Flatiron Institute, Center for Computational Mathematics, New York, USA}
\author{Sonya M. Hanson}
\affiliation{Flatiron Institute, Center for Computational Biology, New York, USA}
\author{Erik H. Thiede}
\affiliation{Department of Chemistry, Cornell University, Ithaca, USA}
\alsoaffiliation{Flatiron Institute, Center for Computational Mathematics, New York, USA}
\email{eht45@cornell.edu}

\title[An \textsf{achemso} demo]
  {Error Breakdown and Sensitivity Analysis of Dynamical Quantities in Markov State Models}

\abbreviations{IR,NMR,UV}

\begin{document}







\begin{abstract}
  Markov state models (MSMs) are widely employed to analyze the kinetics of complex systems. But despite their effectiveness in many applications, MSMs are prone to systematic or statistical errors, often exacerbated by suboptimal hyperparameter choice.
  In this paper, we attempt to understand how these choices affect the error of estimates of mean first-passage times and committors, key quantities in chemical rate theory.  We first evaluate the performance of the recently introduced ``stopped-process estimator''\cite{strahan2021long} that attempts to reduce error caused by choosing a too-large lag time. We then study the effect of statistical errors on Markov state model construction using the condition number, which measures an MSM's sensitivity to perturbation.
  This analysis helps give an intuition into which factors cause an MSM to be more or less sensitive to statistical error.
  Our work highlights the importance of choosing a good sampling measure, the measure from which the initial points are drawn, and has implications for recent work applying a variational principle for evaluating the committor.
\end{abstract}


\section{Introduction}
\label{sec:introduction}

Markov state models (MSMs) offer a powerful framework for understanding the dynamics of chemical and statistical systems. 
Often applied to trajectories from Molecular Dynamics simulations, 
MSMs discretize continuous state spaces into state clusters and count transitions between them to build a reduced model of complex dynamics.  
MSMs have had a successful track record in providing insight for biological systems, having been successfully applied in the study of protein folding~\cite{noe2009folding,voelz2010ntl9}, and more recently to access a better understanding of biological macromolecules related to cancer drug discovery~\cite{hanson2019ddr1,shi2019setd8}, Alzheimer's~\cite{stuchell2023apoe}, and cryptic pocket identification~\cite{bowman2012cryptic}.
Moreover, MSMs can be constructed over large datasets of  molecular dynamics trajectories, helping users to parallelize molecular dynamics simulations~\cite{singh2024massive}.

There has been substantial work on understanding error in MSMs.\cite{sarich2010approximation,prinz2011MSM,djurdjevac2012estimating,schutte2015critical,llamazares2024data,fornace2025approximation}  Most of this work has focused on calculating the error in spectral quantities, where the slowest dynamical modes of the system are approximated using an MSM's eigenfunctions. In this paper, however, we focus on estimating rates, committors, and mean first-passage times (MFPTs).  These quantities have different error behavior than eigenvectors and eigenvalues of the MSM.  For instance, in contrast to the eigenproblem, both the committor and MFPT have biases and errors that increase as the lag time — the interval over which transitions are observed — increases\cite{suarez2016accurate}. These errors are nontrivial, and their behavior under various lag times and sampling probabilities has not been fully understood.

The error with increasing lag time has motivated new approaches to construct Markov State Models.   History-augmented MSMs~\cite{suarez2016accurate,suarez2021what}, for instance, explicitly track the origin of each trajectory in an MSM.  This eliminates bias in estimates of the mean first-passage time.  However, it requires explicit knowledge of a trajectory's past, limiting it's applicability.  Here, we consider the general case, where history information is not present.
Recently, an approach known as the stopped-process estimator was introduced, which mitigates biases from longer lag times without needing history information. The stopped-process estimator\cite{strahan2021long} modifies the system's evolution by stopping trajectories once they hit boundary states, ensuring that important transitions are captured, even at larger lag times. However, the statistical properties of this estimator remain poorly understood. We aim to assess its statistical performance rigorously. 
Furthermore, we analyze the noise sensitivity of MSMs by utilizing the condition number and use this analysis to develop intuition for what factors cause MSMs to be sensitive to noise.
As part of this work, we highlight the crucial role that an MSM's sampling probability plays in the model's estimation error. By adjusting the sampling probability, the distribution of trajectory data can dramatically impact the error behavior of the MSM, particularly for complex systems with intricate dynamical landscapes.

To support this investigation and conduct numerical experiments, we have created a compact, custom framework from scratch, allowing complete control over the algorithms used for MSM construction, error analysis and error diagnostic tools. This framework ensures reproducibility and offers flexibility in testing a variety of MSM estimators with different parameters and sampling strategies. The code repository is available for public access and future use: \url{https://github.com/thiedelab/msm_conditioning}.

The remainder of this paper is structured as follows: in Section \ref{sec:background}, we provide the mathematical background of MSMs and the stopped-process estimator. Section \ref{sec:exp-results} discusses the sources of error in MSMs and presents results of our numerical experiments, highlighting error behavior across different lag times and sampling probabilities. Section \ref{sec:quantifying-statistical-error} introduces the tools for error quantification.  In Section~\ref{sec:discussion} we assess the introduced sensitivity measurement tools and their utility, along with a mathematical connection to variational approaches.  We conclude with a brief discussion of the implications of our findings and potential future work.

\section{Background}
\label{sec:background}

To better guide the reader through the rest of the manuscript we start with the key terms when building an MSM and estimating MFPTs and committors.

\subsection{Markov State Models}\label{ssec:msms}

To construct a an MSM, one first subdivides the phase space into a collection of $K$ disjoint sets $\{\MSet_i\}^K_{i=1}$ known as ``Markov states''.
Typically, this is done by running a clustering algorithm on a large collection of molecular structures.
Next, a molecular simulation or collection of simulation trajectories is divided into a set of time pairs, $(X_0^n, X_\tlag^n)$, where $X_0^n$ is the point in phase space the system occupies at the start of the interval and $X_\tlag^n$ is the point the molecule  occupies after following it for a time $\tlag$.  For concreteness, we assume that this process generates $N$ total time pairs.
We then count the number of time pairs that go from one state to another to build a normalized count matrix with entries
\begin{equation}
    \bar{\Cmat}_{ij}(\tlag) = \frac{1}{N}\sum_{n=1}^{N} \1_{\MSet_i}(X_0^n) \1_{\MSet_j}(X_\tlag^n)
        \label{eq:cmat-mc-approx}
\end{equation}
The function $\1_{\MSet_i}(x)$ is an indicator function,
\begin{equation}
    \1_{\MSet_i}(x) = 
    \begin{cases}
        1, & \text{if } x \text{ is in } \MSet_i, \\
        0, & \text{if } x \text{ is not in } \MSet_i,
    \end{cases}
\end{equation}
and the bar symbol $\bar{\;}$ denotes an average over a finite number of samples.
As the number of time pairs $N$ increases, we expect our normalized count matrix converge to a matrix of correlation functions
\begin{equation}
    \Cmat_{ij}(\tlag) = \iint  \1_{\MSet_j} (y) \1_{\MSet_i} (x) p_\tlag (y | x) \mu(x) dx dy
    \label{eq:true_count_matrix}
\end{equation}
where $p_\tlag(y | x)$ is the probability of the molecule moving from state $x$ to state $y$ in time $t$
and $\mu(x)$ is the probability density from which our initial points $X_0^n$ are drawn. 

Once we have constructed the normalized count matrix, we can 
estimate the probability of seeing a transition from $\MSet_i$ to $\MSet_j$ by constructing the empirical transition matrix
\begin{equation}
    \bar{\Tmat}_{ij}(\tlag)  = \frac{\bar{\Cmat}_{ij}(\tlag)}{\sum_k \bar{\Cmat}_{ik}(\tlag)}.
\label{eq:tmat-mc-approx}
\end{equation}
Similarly, our empirical transition matrix $\bar{\Tmat}_{ij}$ converges to the ``true'' transition matrix
\begin{align}
    \Tmat_{ij}(\tlag) =& \frac{\iint  \1_{\MSet_j} (y) \1_{\MSet_i} (x) p_\tlag (y | x) \mu(x) dx dy}{
        \int \1_{\MSet_i} (x) \mu(x) dx
    } \\
    =&
        \left(C_{ii}(0)\right)^{-1} C_{ij}(\tlag)
    \label{eq:transition_matrix}
\end{align}
where in the last line, we have used the fact that if $0$ time elapses $x=y$, we have that
\begin{equation}
    C_{ij}(0) = \begin{cases}
        \int \1_{S_i}(x) \mu(x) dx &\text{ if } i=j \\
        0 &\text{ otherwise.}
    \end{cases}
    \label{eq:t_0_corr_mat}
\end{equation}

The transition matrix is stochastic, meaning that every element is greater than zero and each row sums to one.
We can interpret the matrix as modeling the dynamics between states: 
element $i,j$ models that chance that the system hops from Markov state $i$ to Markov state $j$ in a given time lag.
Using this model, we can extract key quantities that tell us about the system's kinetics.
For instance, the  eigenfunctions of the transition matrix can be used as approximations of the system's slow modes\cite{noe2013variational,klus2018data}.
Alternatively, we can use the transition matrix to approximate key quantities in chemical rate theory, particularly the the mean first-passage time and the committor\cite{kemeny1969finite, metzner2009transition}.
Assessing the accuracy of these approximations and what causes them to fail is the main goal of this paper.

\subsubsection{Markov State Models as approximators of dynamics}

Both the mean first-passage time and committor are nontrivial quantities to calculate for the system's true dynamics.
Let $\truestopt{A}$ be the first time the system enters $\MSMA$, a set in phase space.
The mean first-passage time at $x$,
a quantity intimately connected to the Kramer's theory of the reaction rate\cite{hanggi1990reaction},
is defined as
the average value of $\truestopt{A}$ over all trajectories that start at $x$:
\begin{equation}
    m(x) = \E \left[ \truestopt{A} \mid X_0 = x \right]
\end{equation}
where $\E$ is the expectation value.  
Another is the committor function:
for two sets in phase space $A$ and $B$ (typically associated with the reactants and products),
the forward committor function $q(x)$ is the probability that a trajectory starting at $x$ will reach $B$ before $A$:
\begin{equation}
    q(x) = \P \left[ \truestopt{B} < \truestopt{A} \mid X_0 = x \right].
\end{equation}
The committor function is the central quantity in transition path theory\cite{metzner2006illustration,vanden2006towards,metzner2009transition}.
The backward committor is defined similarly for reversed time dynamics, and for reversible systems is equivalent to $1 - q(x).$ Here, we consider the forward committor, and refer to it as the committor for simplicity. 

Estimating these quantities for the system's true dynamics in phase space can be difficult.
However, these quantities are easily estimated for a finite-state Markov chain.
Consider a Markov chain $Y_s$ that jumps between $K$ states every $\tlag$ time units, with transition probabilities given by the true MSM transition matrix $\Tmat$. 
Now, let $\MSMA$ and $\MSMB$ be sets of Markov states that cover $A$ and $B$ respectively.
The mean first-passage time for $Y_s$ to enter a collection of Markov states $\MSMA$ is given by
\begin{equation}
    \begin{split}
    m_i =& \sum_{j=1}^K 
    \Tmat_{ij} m_j + \tlag \text{ for } i \in \MSMA^c
    \\
    m_i =& 0 \quad \text{for } i \in \MSMA.
    \label{eq:msm_mfpt} 
    \end{split}
\end{equation}
Here $\MSMA^c$ is the complement of $\MSMA$, namely all of the Markov states that are not in $\MSMA$.
Similarly, the committor function obeys a similar equation,
\begin{equation}
    \begin{split}
    q_i =& \sum_{j=1}^K 
        \Tmat_{ij} q_j \text{ for } i  \in \left(\MSMA \cup \MSMB \right)^c
    \\
    q_i =& 1 \quad \text{for } i \in \MSMB \\
    q_i =& 0 \quad \text{for } i \in \MSMA . 
    \label{eq:msm_committor} 
    \end{split}
\end{equation}

\subsubsection{Solving MSMs using correlation matrices}\label{ssec:msm_to_corr}
Although the transition matrix is more intuitive for formulating the committor and MFPT, for our analysis, it will be more convenient to rewrite
equations~\eqref{eq:msm_mfpt} and~\eqref{eq:msm_committor} using correlation matrices.

Following~\cite{strahan2021long}, we observe that they both the committor and MFPT take the general form
\begin{equation}
    \begin{split}
    u_i =& \sum_{j=1}^K 
        \Tmat_{ij} u_j  + f_i \text{ for } i \in  \MSMD
    \\
    u_i =& g_i  \quad \text{for } i \in \MSMD^c.
    \label{eq:msm_generic} 
    \end{split}
\end{equation}
where $u$ is the unknown function we wish to approximate,  $\MSMD$ is a subset of Markov states that form the ``domain'' on which we solve the problem.
Combining equations and rearranging gives us
\begin{equation}
    \sum_{j \in \MSMD} \left(I - \Tmat \right)_{ij} u_j = f_i  + \sum_{j \in \MSMD^c} T_{ij} g_j
\end{equation}
Multiply both sides by $\int \1_{\MSet_i} \mu(x) d x$, we can rewrite this in terms of the correlation matrix
\begin{equation}
    (\Cmat(t) - C(0))^{\MSMD} u = b,
    \label{eq:generic_linear_eqn}
\end{equation}
where $\Cmat(t)$ is defined as before, $\Cmat(0)$ is the matrix with entries
\begin{equation}
    C_{ij}(0) = \begin{cases}
        \int \1_{\MSet_i}(x)  \mu(x) d x \text{ if } i = j \\
        0 \text{ otherwise,}
    \end{cases}
\end{equation}
with $(C(t) - C(0))^{\MSMD}$ denoting the restriction to just indices in the MSM domain $\MSMD$.
The vector $b$ in~\eqref{eq:generic_linear_eqn} is a vector that depends on the quantity being calculated: for the mean first-passage time,
\begin{equation*}
    b^{\text{mfpt}}_i = - C_{ii}(0),
\end{equation*}
and for the committor, we have
\begin{equation*}
    b^{\text{comm}}_i = - \sum_{j \in B} C_{ij}(t).
\end{equation*}
These suggests a strategy for approximating these quantities for the true dynamics.  
We assume the system's true dynamics are similar to the dynamics of a Markov chain whose transition probabilities are given by the transition matrix in~\eqref{eq:transition_matrix}.
This lets us solve the corresponding version of~\eqref{eq:generic_linear_eqn} with approximations of $C(t)$ and $C(0)$ generated from data.
Finally, for any point in phase space, we approximate the value of the dynamical quantity with the entry in $u$ for the corresponding Markov state.

Importantly, this can be a much more computationally effective way of constructing rate estimates.   Constructing a Markov state model only requires trajectories of time length $t$.
Since this may be a much shorter time than typical values of $\truestopt{A}$ or $\truestopt{B}$,
in theory MSMs give us a way to estimate long-timescale quantities using only short trajectories.

\subsection{Operator equations for the true dynamics}
To evaluate the accuracy of our approximations to the mean first-passage time and committor function,
we first seek to understand what equations~\eqref{eq:msm_mfpt} and \eqref{eq:msm_committor} are approximating.
Assume that the dynamics of the system is given by a discrete-time Markov process $X_s$ that takes a step every $\tstep$ time units according to a time-invariant transition probability density. 
We can define an operator $\Koop{t}$ that acts on functions $f(x)$ on phase space, and evolves them according to the dynamics of the system.
\begin{align*}
    \Koop{t} f(x) =& \E \left[ f(X_t) \mid X_0 = x \right] \\
    =& \int f(y) p_t(y | x) dy,
\end{align*}
where $p_t(y | x)$ is the probability of the system moving from $x$ to $y$ in time $t$ as before.
This operator, known as the \emph{stochastic Koopman} operator or \emph{transition operator}, is the continuous-space generalization of the transition matrix~\cite{klus2018data}. 
Indeed, comparing with~\eqref{eq:true_count_matrix}, we see that we can write the correlation matrix as
\begin{equation}
    C_{ij}(t) = \int \1_{S_i}(x) \Koop{t} \1_{S_j}(x) \mu(x) dx
    \label{eq:koopman_to_correlation_matrix}
\end{equation}
and the transition matrix as,
\begin{equation}
    T_{ij}(t) = \frac{
        \int \1_{S_i}(x) \Koop{t} \1_{S_j}(x) \mu(x) dx
    }{
        \int \1_{S_i}(x)  \mu(x) dx
    }
    \label{eq:koopman_to_transition_matrix}
\end{equation}

\subsubsection{Connecting the true and MSM dynamics}

With the transition operator, we can also generalize~\eqref{eq:msm_mfpt} and~\eqref{eq:msm_committor} to the true mean first-passage time and committor.
The true mean first-passage time is the solution to the following operator equation:
\begin{equation}
    \begin{split}
        m(x) &= \Koop{\tstep} m(x) + \tstep \text{ for } x \in  A \\
    m(x) &= 0 \quad \text{for } x \in  A.
    \label{eq:main-equation-mfpt}
    \end{split}
\end{equation}
Similarly, the true committor function solves
\begin{equation}
    \begin{split}
    q(x) &= \Koop{\tstep} q(x) \text{ for } x \notin A \cup B \\
    q(x) &= 1 \quad \text{for } x \in B \\
    q(x) &= 0 \quad \text{for } x \in A.
    \label{eq:main-equation-committor}
    \end{split}
\end{equation}
We note that these equations only hold when using the Koopman operator for a single step, $\Koop{\tstep}$.  If we replace it with the Koopman operator at a different time, $\Koop{t}$, these equations are at best approximately correct.

\subsubsection{Rederiving MSM equations from the Koopman operator}\label{sssec:dga}

In fact, we can derive Equations~\eqref{eq:msm_mfpt} and~\eqref{eq:msm_committor} by projecting 
the transition operator
onto indicator functions on the MSM sets.\cite{thiede2019galerkin}.
For concreteness, we focus on deriving~\eqref{eq:msm_committor} from~\eqref{eq:main-equation-committor}.
The derivation of~\eqref{eq:msm_mfpt} from~\eqref{eq:main-equation-mfpt} is similar.

We first approximate the committor function to be piecewise constant on each of our Markov states:
\begin{equation}
    q(x) \approx \sum_{j=1}^K q_j \1_{S_j}(x).
    \label{eq:committor_sum_of_indicators}
\end{equation}
Each Markov state is assumed to be wholly in either $A$, $B$, or neither.
This means that for points in either $A$ or $B$, this approximation is exact: we just set $q_j$ to $0$ or $1$ for the corresponding Markov state.

However, for the other points~\eqref{eq:committor_sum_of_indicators} is indeed an approximation, and~\eqref{eq:main-equation-committor} will not hold.  Consequently, we require a weaker condition.
Multiplying by $\1_{S_i}(x)$ and integrating over an arbitrary probability density $\mu(x)$, we merely require that 
\begin{equation*}
    \int\! \1_{S_i}(x) q(x) \mu(x) dx = \int \! \1_{S_i}(x) \Koop{\tstep} q(x) \mu(x) dx
\end{equation*}
for every Markov state $S_i$ that is not contained in $A$ or $B$.
Substituting in our approximation~\eqref{eq:committor_sum_of_indicators} to the left and right-hand sides above, we have
\begin{align*}
    \sum_{j=1}^K  q_j & \int \1_{S_j}(x)  \1_{S_i}(x) \mu(x) {d}x \\
    &=
    \sum_{j=1}^K q_j \int \1_{S_i}(x)  \Koop{\tstep} \1_{S_j}(x) \mu(x) {d}x 
\end{align*}
Applying~\eqref{eq:koopman_to_correlation_matrix} and~\eqref{eq:t_0_corr_mat}  and rearranging gives us the matrix
equation
\begin{align*}
    \sum_{j\notin \MSMA \cup \MSMB} \left(C_{ij}(\tlag) - C_{ij}(0)\right) q_j =& -\sum_{j \in \MSMB } q_j C_{ij}(0) \\
    =& -\sum_{j \in \MSMB} C_{ij}(0),
\end{align*}
the equation we presented in Subsection~\ref{ssec:msm_to_corr} for a lag time of $\tstep$.

\subsubsection{Increasing the lag time}

This projection can introduce considerable error.
One historical strategy for reducing the error is to build Markov state models at a larger time interval $t> \tstep$, referred to as the \textit{lag time}.\cite{swope2004describing,husic2018markov}
Formally, this corresponds to approximating Equations~\eqref{eq:main-equation-mfpt} and~\eqref{eq:main-equation-committor} 
by replacing $\tstep$ with a longer time lag $\tlag$, and then approximating this biased operator equation with a Markov state model.
The hope is that at large lag times, rapidly mixing dynamics will have converged, making $\Koop{t}$ easier to approximate.
However, the \emph{true} mean first-passage time and committor functions do not satisfy these equations:
by increasing the lag time, we introduce a systematic bias into our estimates.
For very large lag times, our estimates may tell us nothing about the system's actual kinetics.  As an extreme case, for very long lag times the system's current state may be completely decorrelated from its starting point.  At that point, the system's current state has nothing to do with kinetics but is rather determined by equilibrium statistical mechanics. 

\subsection{The stopped process estimator}

The cause for this bias is simply that we miss important events that occur on shorter timescales.
It is possible that between time $0$ and time $\tlag$ the system has entered and exited states $A$ or $B$ multiple times.  However, $\Koop{\tlag}$ only accounts for the state of the system at time $\tlag$.  Any events that occur before $\tlag$ are ignored, inflating our estimate of the mean first-passage time\cite{suarez2016accurate} and biasing our value of the committor.

A simple fix was proposed in ref.~\citenum{strahan2021long}.  
Similar to the approach used to construct core-set MSMs,\cite{buchete2008coarse} or certain formulations of Milestoning,\cite{vanden2009markovian} as soon as the system enters a boundary state, we record the state of the system and stop any further movement.

\begin{figure*}[h!]
    \centering
    \includegraphics[width=0.9\linewidth]{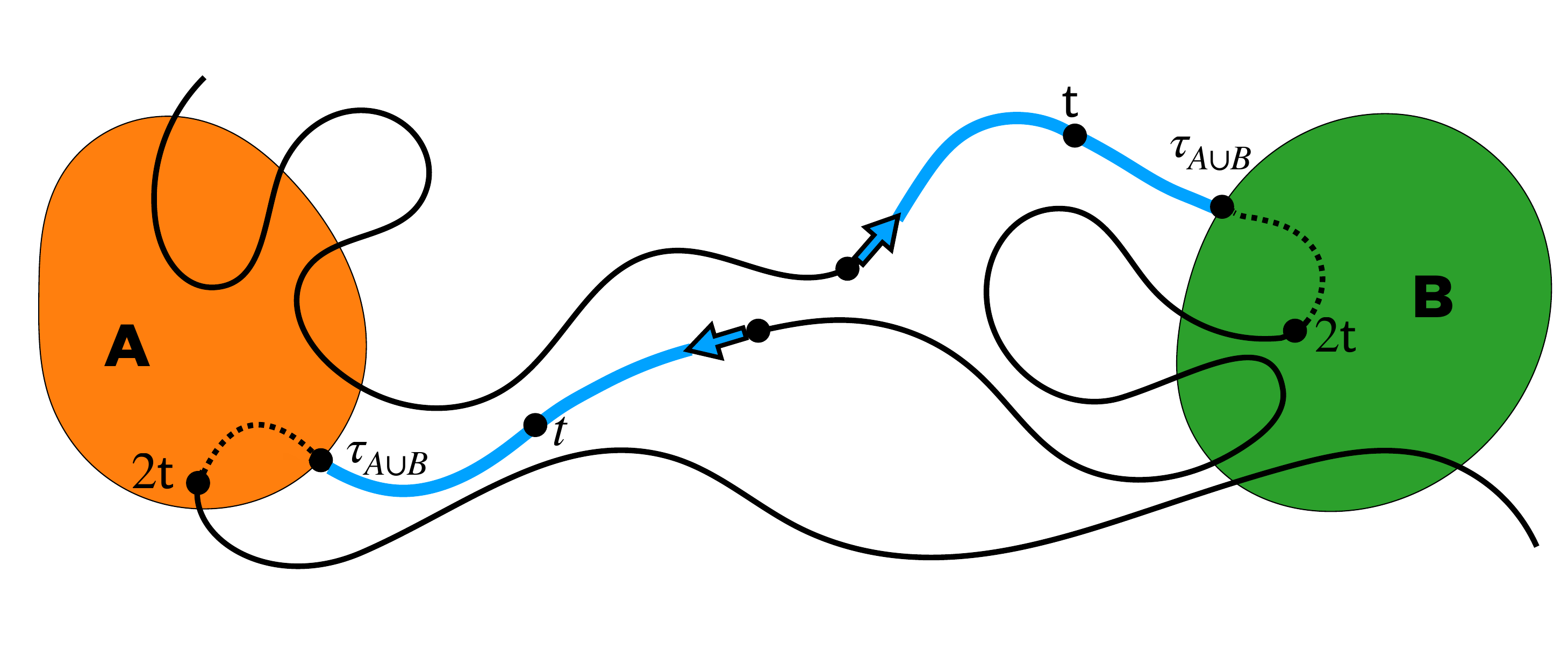}
    \caption{A trajectory in a phase space passing through the reactant set $A$ and the product set $B$. For a stopped Markov process, the trajectory is modified such that once it enters the boundary states (either in $A$ or $B$), the dynamics do not continue.
    Consequently the probability of leaving these states is zero, a fact that is reflected in the transition matrix $T_{ij}$ of the stopped Markov process.
    }
    \label{fig:tpt-explanatory}
\end{figure*}

We refer to this modified Markov chain as the \textit{stopped process} and depict it in Figure~\ref{fig:tpt-explanatory}.
We denote the set of boundary states as $D^c$, and define the stopping time $\truestopt{D^c}$ as the first time the system enters $D^c$.
Then letting  
\begin{equation}
    \tlag \wedge \truestopt{D^c} = \min\{\tlag, \truestopt{D^c}\},
\end{equation}
we can define the Koopman operator for the stopped process as the operator that evolves a function $f(x)$ according to the dynamics of the stopped process:
\begin{equation}
    \SKoop{\tlag} f(x) = \E \left[ f( X_{\tlag \wedge \truestopt{D^c}}) \mid X_0 = x \right].
\end{equation}
Using the Koopman operator for the stopped process, we can define an equation for the committor that holds for any lag time.
\begin{equation}
    \begin{split}
        q(x) &= \SKoop{\tlag} q(x) \text{ for } x \notin A \cup B \\
    q(x) &= 1 \quad \text{for } x \in B \\
    q(x) &= 0 \quad \text{for } x \in  A.
    \label{eq:main-equation-committor-stopped}
    \end{split}
\end{equation}
Since the dynamics have been stopped as soon as the system enters $A$ or $B$, we do not miss any events where the system enters $A$ or $B$.  Consequently, considering a longer time interval $t$ does not introduce bias.
Similarly, the mean first-passage time for the stopped process satisfies
\begin{equation}
    \begin{split}
    m(x) &= \SKoop{\tlag} m(x) + \E[\tlag \wedge \truestopt{D^c}] \text{ for } x \notin A \\
    m(x) &= 0 \quad \text{for } x \in A.
    \label{eq:main-equation-mfpt-stopped}
    \end{split}
\end{equation}

We can also construct correlation matrices and transition matrices for the stopped process: the correlation matrices for stopped process are given by
\begin{align}
        \SCmat_{ij}(\tlag) =& \int \1_{\MSet_i}(x) \Koop{\tlag}_{D^c} \1_{\MSet_j}(x) \mu(x) dx 
        \label{eq:stopped_count}
        \\
    \SCmat_i(0) =& \int \1_{\MSet_i}(x) \mu(x) dx,\label{eq:stat_stopped_count}
\end{align}
and the corresponding transition matrix is
\begin{equation}
    \STmat_{ij}(\tlag) = \left( \SCmat_{ij}(0) \right)^{-1}\SCmat_{ij}(\tlag) \label{eq:stopped_koopman_tmat}
\end{equation}

Moreover,
we can use the same derivation described in Subsubsection~\ref{sssec:dga}.
For the mean first-passage time, we solve
\begin{equation}
    \label{eq:main-equation_stopped-projected}
    \sum_{j \notin A}^K  \left(\SCmat_{ij}(\tlag)  - \SCmat_{ij}(0) \right)m_j = h_i
\end{equation}
where we have defined 
\begin{equation}
    h_i = \int \1_{\MSet_i}(x) \E[\tlag \wedge \truestopt{D^c}] \mu(x) dx.
\end{equation}
and $m_i = 0$ for all states in $A$.
Similarly, we can express a system of equations for the committor function.
\begin{align}
    \label{eq:main-equation_stopped-committor}
    \sum_{j\notin \MSMA \cup \MSMB}^K \left(\SCmat_{ij}(\tlag) - \SCmat_{ij}(0) \right)q_j = -\sum_{j\in \MSMB} \SCmat_{ij}(\tlag)
\end{align}
To differentiate between the correlation matrices presented in Subsection~\ref{ssec:msms} and the ones presented here, we will refer to the count matrix and transition matrix presented here as the ``stopped process'' estimates, and the matrices presented in Subsection~\ref{ssec:msms} as the ``naive'' estimates. 
We stress that this is not a value judgment: rather, we choose this nomenclature as the estimate is ``naive'' to what a trajectory does between time $0$ and time $\tlag$.

\begin{figure*}[h!]
    \centering
    \begin{subfigure}[b]{0.49\linewidth}
        \centering
        \includegraphics[width=\linewidth]{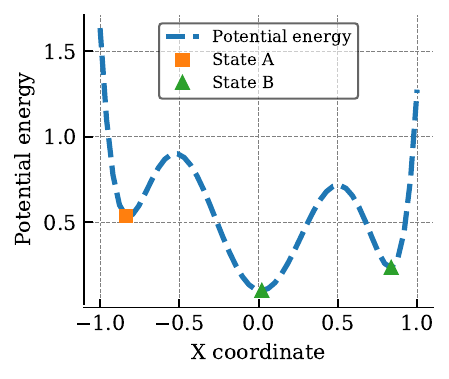}
        \caption{Triple-well potential}
        \label{fig:triple-well-potential}
    \end{subfigure}
    \hfill
    \begin{subfigure}[b]{0.49\linewidth}
        \centering
        \includegraphics[width=\linewidth]{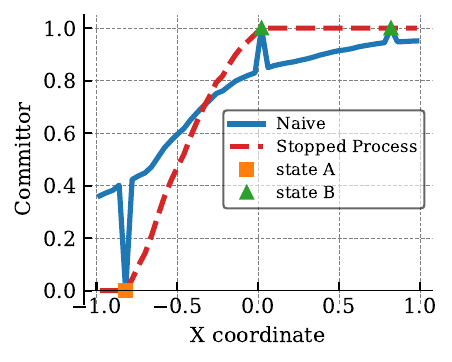}
        \caption{Committor on triple-well}
        \label{fig:committor-on-triple-well-illustrative}
    \end{subfigure}
    \caption{Comparing naive vs stopped process committors. For large enough lag times trajectory can hop over green states and thus miss B hitting events, which causes the naive estimator to have increasing error bias. However, since the system is one-dimensional this is unphysical: any continuous trajectory starting at any point between the green states cannot reach the orange state before hitting one of the green states.}
    \label{fig:triple-well-whole-figure}
\end{figure*}

To help demonstrate the source of this error, we construct an one-dimensional example, depicted in Figure~\ref{fig:triple-well-whole-figure}.  Our example consists of a triple well system.  State A corresponds to the minimum of the bottom left well.  State B, in contrast, is split: part is in the middle well, and the other part is in right well.

With this construction, the committor is zero left of state A.  As we pass state A, the committor increases as we get closer to the middle well, reaching one at its center.  Between the second and third well, the committor is 1 everywhere: it is impossible to go from the third well to state A without first hitting the second well.

In Figure~\ref{fig:triple-well-whole-figure}, we calculate the committor using both the naive and stopped process estimators.  The stopped process estimator correctly has a committor of 1 between the second and third well.  However, the committor for the naive process actually decreases below zero.  This is because the naive estimator doesn't count  events where the system enters the second well and then passes over it.  Consequently, in the resulting MSM the system  can ``hop over'' the center well and head straight to state A, leading to an erroneously low committor.

\section{Demonstrating sources of error}
\label{sec:exp-results}

In this section, we address the accumulation of error in estimating dynamical quantities such as the committor and MFPT within the MSM framework. We focus on two central questions: (1) How does the choice of \textbf{lag time} affect the accuracy of dynamical quantity estimates? (2) How does the \textbf{sampling probability} influence the stability and precision of these estimates? Through a series of numerical experiments, we aim to explore the relationship between these factors and provide insights for optimizing MSM-based estimations.

An experimental MSM is constructed based on a set of trajectories. To assess the impact of the trajectory dataset on the accuracy of correlation matrix estimations (see equation \ref{eq:cmat-mc-approx}), we consider two aspects: (a) the total number of individual \textit{transition events} $(X_0, X_{\tlag})$ available, and (b) how these transitions are distributed across the phase space, which is influenced by the sampling probability.

While the number and distribution of state-to-state transitions are key factors in determining the accuracy of dynamical quantity estimates, some errors remain even when MSMs are estimated from infinitely long trajectories. This leads to a distinction between two types of dynamical quantity errors based on whether they persist in the limit of infinite data:

\begin{enumerate}
    \item \textbf{Approximation error}: 
    This is the error due to the model assumptions and hyperparameters, particularly the approximation of the transition operator in~\eqref{eq:koopman_to_transition_matrix} by the projection onto indicator functions. Stated differently, approximation error is the error still present even if all Monte Carlo averages such as~\eqref{eq:tmat-mc-approx} were exact, i.e would persist even as the number of transition samples increases.
    \item \textbf{Estimation error}:
    This error is the error from finite length of trajectories and finite number of transition events, such as the monte carlo averaging in~\eqref{eq:tmat-mc-approx}. 
    Estimation error is statistical in nature and diminishes as the amount of data increases, converging to zero in the infinite data regime. 
\end{enumerate}
 
\subsection{Behavior of the schemes with varying lag time}

\begin{figure*}
    \centering
    \begin{subfigure}[b]{0.49\linewidth}
        \centering
        \includegraphics[width=\linewidth]{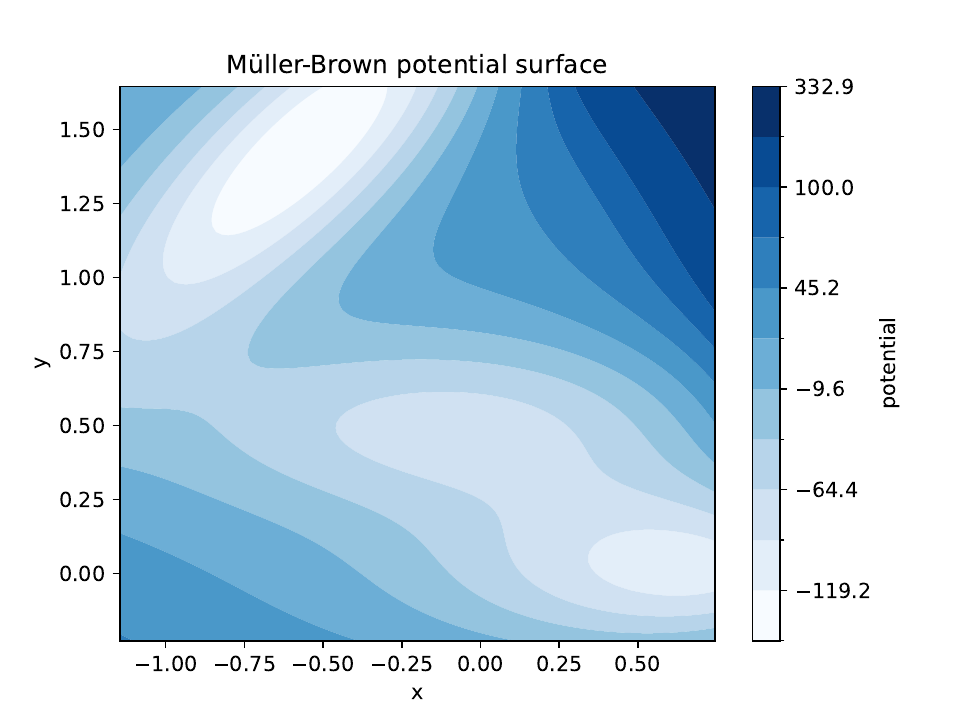}
        \caption{}
        \label{fig:mb-potential-surface}
    \end{subfigure}
    \hfill
    \begin{subfigure}[b]{0.49\linewidth}
        \centering
        \includegraphics[width=\linewidth]{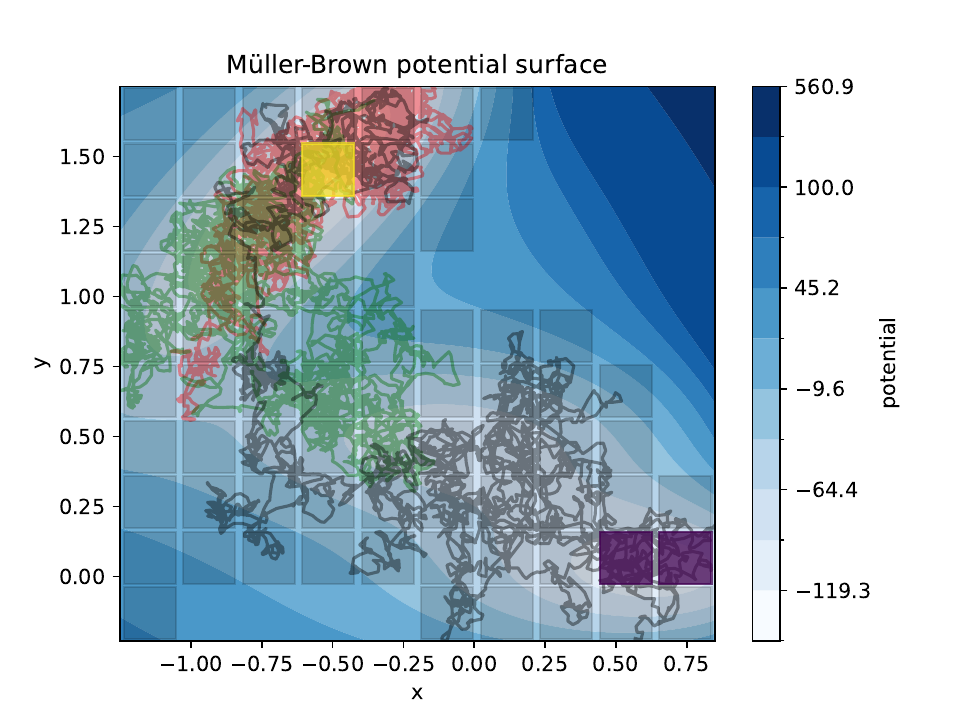}
        \caption{}
        \label{fig:mb-traj-and-clusters}
    \end{subfigure}
    \caption{(a) Müller-Brown potential surface. (b) Grid of Markov states used to construct the MSMs, plotting along with select trajectories an example dataset.  Yellow clusters represent product $\MSMB$ states, purple clusters represent reactant $\MSMA$ states. Gray squares represent domain clusters.}
    \label{fig:mb-depiction}
\end{figure*}

To begin, we explore how the choice of lag time affects both sources of error separately. To begin, we study Brownian dynamics on the M{\"u}ller-Brown potential surface, a classic model system for chemical kinetics experiments\cite{muller1979location}.
We contruct an MSM using a grid of states on the two-dimensional surface. 
We depict the potential energy surface for the M{\"u}ller-Brown potential in Figure~\ref{fig:mb-depiction}, along with a representative set of trajectories used to construct the MSM and the definitions of the states.
Further experimental details are provided in the Supplement in Subsection~\ref{ssec:hyperparams}.
Considering a two-dimensional potential allows us to calculate  an accurate grid-based approximation of the system's true Koopman operator using the scheme presented in \citet{thiede2019galerkin} and calculate accurate reference values for the mean first-passage time and committor.

\begin{figure*}
    \centering
    \begin{subfigure}[b]{0.49\linewidth}
        \centering
        \includegraphics[width=\linewidth]{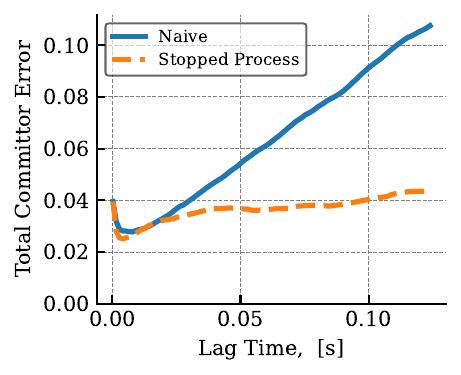}
        \caption{Total Error}
        \label{fig:real-committor-error-of-lag-time}
    \end{subfigure}
    \hfill
        \begin{subfigure}[b]{0.49\linewidth}
        \centering
        \includegraphics[width=\linewidth]{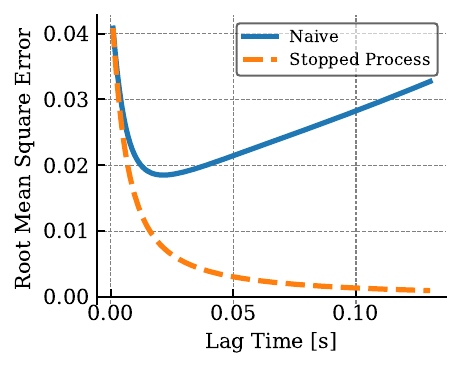}
        \caption{Approximation Error}
        \label{fig:committor-errors}
    \end{subfigure}
    \caption{Approximation and total error in the committor as a function of lag time for Müller-Brown system.
   }
    \label{fig:committor-and-mfpt-inf-data-errors}
\end{figure*}

In Figure~\ref{fig:real-committor-error-of-lag-time}, we show the error in the committor estimate calculated for both the stopped-process and naive estimators.
As the lag time increases, the error from the naive estimator continues to rise, whereas the error from the stopped-process estimator plateaus.
To understand the source of this error, we use our grid-based approximation of the Koopman operator to approximate the matrix entries in~\eqref{eq:koopman_to_transition_matrix} and~\eqref{eq:stopped_koopman_tmat}  in the absence of estimation error.
We depict the resulting approximation error in Figure~\ref{fig:committor-and-mfpt-inf-data-errors}.
For the naive estimator, the approximation error increases as lag time grows due to the omission of boundary hitting events.
This growing bias results from the failure to capture transitions that occur between lag time intervals and hit boundary states, leading to inaccuracies in the estimation of dynamical quantities.
In contrast, the stopped process goes to zero approximation error as the lag time increases.  This is because, as the lag time goes to infinity, the stopped process estimator recovers a direct approximation of the committor or mean first-passage time by counting hitting events.
Consequently, the remaining error in the stopped process estimator in Figure~\ref{fig:real-committor-error-of-lag-time} is estimation error:
even though the stopped process has small bias at large lag times, the statistical error in the estimate nevertheless persists.  

\subsection{Behavior of the schemes with varying sampling probabilities}
\label{sec:sampling-measure-experiments}

To understand the role of estimation error on MSM accuracy, we focus our attention on the sampling measure.  The sampling measure $\mu$ encodes how many trajectories start in each Markov state.  Consequently, it governs how well each Markov state will be sampled  . Careful attention must be paid to the choice of sampling strategy, as it can severely impact the accuracy of the final estimates. 

To illustrate the impact of the sampling probability, we considered a random walk on a one-dimensional double-well potential system, with states in both wells.  
To test the effect of the sampling measure, we interpolate between the Boltzmann distribution and the uniform distribution over our double-well.
We construct a family of sampling measures by defining a density $\mu_{\mixprm}(x) = (1 - \mixprm)  \mu_{\text{Boltzmann}}(x) + \mixprm \mu_{\text{uniform}}(x)$.
Here, we refer to $\mixprm$ as the mixing parameter which controls a smooth linear transition from Boltzmann to uniform sampling probability functions. 

\begin{figure*}[h!]
    \centering
    \begin{subfigure}[b]{0.49\linewidth}
        \centering
        \includegraphics[width=\linewidth]{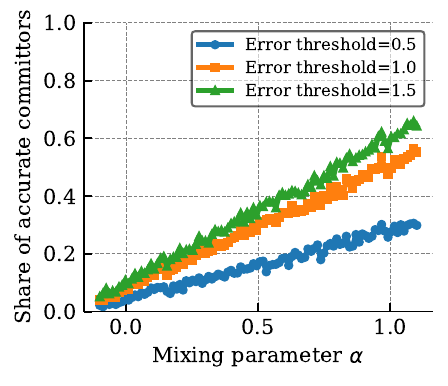}
        \caption{}
        \label{fig:committor-error-of-sampling-strategy}
    \end{subfigure}
    \hfill
    \begin{subfigure}[b]{0.49\linewidth}
        \centering
        \includegraphics[width=\linewidth]{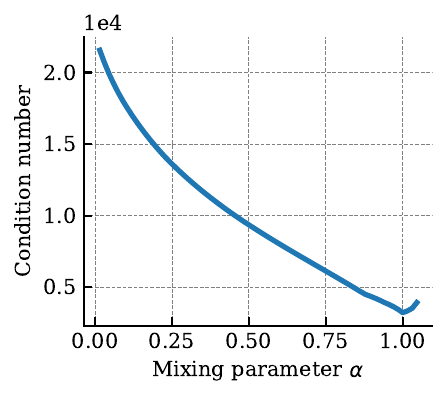}
        \caption{}
        \label{fig:cond-num-of-sampling-measure-homotopy}
    \end{subfigure}
    \caption{Demonstration of how sample strategy choice can affect the error in MSM committor estimates for the double-well system. \textbf{\ref{fig:committor-error-of-sampling-strategy}}: Share of accurate committors as a function of sampling probability temperature. \textbf{\ref{fig:cond-num-of-sampling-measure-homotopy}}: Condition number of the committor equation coefficient matrix as a function linear mixing parameter $\mixprm$ between the Boltzmann and uniform distribution ($\mixprm=0$ corresponds to Boltzmann, $\mixprm=1$ corresponds to uniform).
    }
    \label{fig:experiments-with-sampling-measure}
\end{figure*}

For each value of $\mixprm$, we then sampling 1000 initial points from the sampling measure and construct a Markov state model.  We repeat this process 1000 times and calculate the committor for each replicate.
Importantly, if the sampling measure is poor, the committor calculation can fail because the Markov state model is disconnected or does not have sufficient samples in one state.  Consequently, we introduce a new measure of error that is tolerant to this failure.  For each committor, if the calculation succeeds we evaluate its root-mean-square error, 
\begin{equation*}
     \sqrt{\sum_i (\tilde{q}_i - q^\text{ref}_i)^2}.
\end{equation*}
Otherwise, if the calculation fails, we say the error is infinite.
Then, we calculate the share of committors with an error less than a given tolerance.
We plot this share as a function of $\alpha$ in Figure~\ref{fig:committor-error-of-sampling-strategy}. 
Notably, sampling probabilities that are close to uniform sampling perform better in estimating the correlation matrices, resulting in a more accurate committor function estimation. 
This occurs because at low values of $\mixprm$ we are sampling close to the Boltzmann distribution.  Consequently, samples are drawn from the metastable low-free energy wells, but not from other important Markov states such as states located at transitions.

\section{Understanding MSM estimation error}
\label{sec:quantifying-statistical-error}
Our results highlight the importance of understanding estimation error in MSMs.
Good tools for constructing estimates of statistical error for MSMs already exist.
If enough data is available, errors can be directly quantified using bootstrapped error estimates\cite{chodera2007automatic,bowman2009using} to quantify the estimation error.
Alternatively, asymptotic analysis can be used to derive central limit theorems that can be used to estimate the variance in the large sample limit\cite{singhal2005error,cheikhi2023statistical,cheng2024surprising}.  

Here, our goal is to understand what factors cause an MSM to have large estimation errors.
To this end, we derive simple mathematical relations that describe the error in MSM estimates of rates and committors
using the condition number, a common way of measuring sensitivity to perturbation in numerical linear algebra.  We first review the mathematical background of the condition number before applying it to understanding errors in MSMs.

\subsection{Assessing the error using condition numbers}

To understand the effect of estimation error on an MSM estimate, we must assess both the size of the error as well as the magnitude of its effect on the final estimates.
To do this, we need a way to quantify ``size'' of a vector or matrix.  
This is given by the notion of a \textit{norm}, which is a generalization of the absolute value\cite{deif2012sensitivity}.
For a vector $v$, the $p$-norm is defined as
\begin{equation}
    \|v\|_p = \left( \sum_i |v_i|^p \right)^{1/p}.
\end{equation}
where $p$ is a positive real number.  Different norms give different notions of size: The $2$-norm, for instance, gives the Euclidean length of a vector.  We will instead primarily use the $1$-norm, defined as 
\begin{equation}
        \|v\|_1 = \sum_i |v_i|
\end{equation}
as it often leads to convenient probabilistic interpretations.

Vector norms induce a notion of the size for matrices, aptly called the \textit{induced matrix norm}.  For a matrix $A$, the induced matrix $p$-norm is defined as\cite{belitskii2013matrix}
\begin{equation}
    \|A\|_p = \max_{\|v\|_p =1 }  \|A v \|_p 
\end{equation}
In words, the induced norm of a matrix shows how much the matrix can stretch a vector with norm 1.
For arbitrary $p$ there is no simple formula for the induced matrix norm, but for specific induced norms there are simple interpretations.  In particular, the induced $1$-norm of a matrix obeys
\begin{equation*}
    \|A\|_1 = \max_j \sum_i |A_{ij}|.
\end{equation*}

\subsubsection{Condition numbers}\label{sssec:condition_numbers}

Equipped with the notion of a matrix norm, we can present a well-known result from numerical linear algebra\cite{deif2012sensitivity}.
\begin{theorem}
\label{theorem:bound-based-on-cond-num}
Let $(A + E)y = b + e_b$ be a perturbed version of the equation $A x = b$
and $\| \cdot \|$ denote an induced matrix norm.
Furthermore, let $\|A^{-1}\| \|E\| < 1$. Then $A + E$ is nonsingular and
\begin{equation}
   \frac{\|x - y\|}{\|x\|}\leq \kappa(A) \frac{1 }{1\! -\! \|A^{-1}\| \|E\|} \left( \frac{\|E\|}{\|A\|}\! +\! \frac{\|e_b\|}{\|b\|} \right)
   \label{eq:condition_number_bound}
\end{equation}
where $\kappa(A)$ is the condition number of $A$, defined as
\begin{equation}
    \kappa(A) = \|A\| \|A^{-1}\|. 
\end{equation}
\end{theorem}
For a proof, we refer to ref.~\citenum{deif2012sensitivity}.
This theorem motivates the condition number as a measure of how sensitive the solution to a linear system is to errors in the matrix and right-hand side.
A large condition number suggests that a small error matrix $E$ may nevertheless lead to very large 
errors in estimating $x$.
Conversely, a small condition number means that estimates of $x$ are generally error-robust.

Some care is needed when interpreting the condition number.
First, we note that the right-hand-side of~\eqref{eq:condition_number_bound} is a bound, not an equality. 
Often this bound is very loose, and the actual error is much smaller.
Moreover, different norms can lead to different condition numbers,
complicating interpretation.
Nevertheless, the condition number remains a useful tool for judging the sensitivity of a problem, as well as for determining the source of sensitivity to errors.

\subsection{The condition number as a sensitivity metric}\label{ssec:condition_number_for_sensitivity}

We can use this approach to assess how sensitive MSMs are to estimation error.
For both the naive and the stopped process estimator, our estimates of dynamical quantity functions can be rewritten as the solution to a linear system of the following form:
\begin{equation}
    \left(\bar{\Cmat}({\tlag}) - \bar{\Cmat}(0)\right) u = \bar{r}.
    \label{eq:generic_msm_equation}
\end{equation}
where $\bar{r}$ depends on the specific dynamical quantity being estimated and $\bar{\Cmat}$ is the naive or stopped count matrix, as appropriate.  
Comparing with Theorem~\ref{theorem:bound-based-on-cond-num}, we observe that the condition number for our estimates is given by
\begin{equation}
    \kappa(\tlag)\! =\! \big\|\SCmat(\tlag)\! -\! \SCmat(0)    \big \|_p \big\|\!\left(\SCmat(\tlag) \!- \!\SCmat(0) \right)^{-1} \! \big\|_p
    \label{eq:rewritten_condition_number}
\end{equation}
In general, this leads to a conservative estimate of the sensitivity, and the bound in Theorem~\ref{theorem:bound-based-on-cond-num} can overestimate the error by an order of magnitude or more.
To demonstrate this, in the Supplement we calculate the 
bound computed using the condition number in~\eqref{eq:condition_number_bound}
with the true error, along with another bound presented in~\citet{cheng2024surprising}.  Both bounds are much larger than the actual error in the committor estimate.  However, studying the condition number can nevertheless give insights into how noise-sensitive an MSM is.

Different choices of $p$ in the condition number can lead to different error bounds. 
Here, we focus on the $1$-norm, as its simple form allows us to derive interpretable error estimates..
Let $Y_t$ be a Markov chain whose transition probabilities are given by the transition matrix $T$.
We stress that this is \emph{not} the same as the underlying dynamics of the molecular system.
Rather, it is the dynamics the system \emph{would have} if the MSM were a perfect representation of the system.
At time $0$, $Y_0$ is distributed according the vector $\mu_i = \int_{S_i} \mu(x) \text{d}x$.  
With this assumption, both terms in ~\eqref{eq:rewritten_condition_number} have probabilistic interpretations.  We give derivations in subsection~\ref{ssec:cnumber_derivations} of the Supporting Information.

The first term in the condition number can be interpreted as a sum of two probabilities,
\begin{align}
 &\left\| \SCmat(\tlag) - \SCmat(0) \right\|_1 \! 
  \nonumber \\
 &\quad =\max_j \left[ \P(\text{enter }j) + \P(\text{leave }j) \right],
    \label{eq:interpretation_term_1}
\end{align}
where we have defined
\begin{align*}
    \P(\text{enter }j) =& \P\! 
            \left(
                Y_{0}\! \in D,
                Y_{0}\! \neq j, 
                Y_{\tlag}\! = j
            \right) 
            \\
    \P(\text{leave }j)  =& 
    \P\!\left(
                Y_{0}\! = j, 
                Y_{\tlag}\! \neq j
            \right) .
\end{align*}
These two terms correspond to probabilities of observing trajectories enter or leave state $j$:  The first term gives the probability that we observe a trajectory that enters state $j$ after a single step of the MSM dynamics and that it started  in a different state in the domain, and the second term gives the probability that we observe a trajectory that starts in $j$ and then takes a step into a new state in the domain. 

In general, we expect this term to stay reasonably controlled: as the sum of two probabilities of mutually exclusive events, it is trivial to see that it is bounded by $1$.
However, the second term can grow to considerable size.  In the appendix we show that it can be rewritten as
\begin{equation}
\begin{split}
    &\left\| \left(\SCmat(\tlag) - \SCmat(0) \right)^{-1} \right\|_1\! \\
    &\quad = \!\max_j 
    \left[\frac{ \sum_i \E[\sum_{l=0}^\infty \1_j(Y_{lt}) | Y_0 =i]}{\mu_j} \right]
    \label{eq:interpretation_term_2}
    \end{split}
\end{equation}
The expectation in the numerator of the fraction can be interpreted as follows.  For every state $i$, we initialize in $i$ and calculate average number of times that the MSM Markov process enters state $j$ before exiting the domain.
The denominator is simply the sampling measure.  

This fraction is large if (a) the dynamics of $Y_s$ spend considerable time in a state, and (b) few samples are initialized in that state.
For instance, if there are any metastable states in the domain, this term can be particularly large.  At its most extreme, one could consider a disconnected state in the domain, which $Y_s$ never leaves.  Such a state would make the condition number infinity.

\subsection{Heuristics for the error in MSM estimates}

By considering the condition number in specific limits, we can gain insights into the factors that cause MSMs to have large or small staatistical error.

\subsubsection{Effect of the sampling measure}

Our results in Section~\ref{sec:sampling-measure-experiments} suggest that the choice of sampling measure can have a significant impact on the accuracy of MSM estimates. Sampling 
from the equilibrium measure can lead to very large errors.
This is reflected in the condition number: Equation~\eqref{eq:interpretation_term_2} has a $1 / \mu_j$ factor, multiplied by a prefactor that increases the more state $j$ is visited by the system as it leaves the domain.
If initial points are sampled according to the Boltzmann measure, then $1 / \mu_j$ becomes exponentially large, causing both the condition number and the bound on the asymptotic variance to grow quickly.
Consequently, building MSMs in a data-efficient manner may require sampling initial points from a \emph{nonequilibrium} measure. 
This motivates the use of existing approaches that attempt to initialize trajectories out of equilibrium.  For instance, practitioners may choose initial points from simulations run at temperatures higher than the target temperature\cite{bowman2010enhanced} or adaptively focus sampling on regions believed to be important\cite{wan2020adaptive}.  This leads to more even sampling across Markov states, reducing the size $1 / \mu_j$ and leading to better estimates.

\subsubsection{Behavior of the stopped process estimator at large lag times}

We next turn our attention to the behavior of the stopped process estimator at large lag times.
Since, for large $\tlag$, the probability of the underlying process leaving the domain in time $\tlag$ approaches one,
we expect that
\begin{align*}
    \P\! 
    \left(
        Y_{0}\! \neq j, 
        Y_{\tlag}\! = j
    \right) & \xrightarrow[]{\tlag\to\infty} 0
    \\
    \P\! \left( Y_{0}\! = j, Y_{\tlag}\! \neq j \right) & \xrightarrow[]{\tlag\to\infty} \mu_j
    \\
    \E[\sum_{t=0}^\infty \1_j(Y_t) | Y_0 =i] 
    & \xrightarrow[]{\tlag\to\infty} 
    \begin{cases}
        1 \text{ if } i=j \\
        0 \text{ otherwise.}
    \end{cases}
\end{align*}
As a result, our condition number reduces to  
\begin{equation}
    \lim_{\tlag\to\infty}\kappa(\tlag)
        = 
        \left(\max_j \mu_j\right)
        \left(
            \max_j \frac{1}{\mu_j}
        \right).
\end{equation}
Looking at this expression, we see that it has no dependence on the dynamics of the system but instead depends only on the sampling measure.
This is explained by our intuition that the stopped process reverts to the direct counting estimate at long lag times.
Since we are directly counting exit events from the domain, the error is controlled simply by the number of exit events each state sees.  This is given exactly by the sampling measure.

\subsubsection{Comparing the stopped process and the naive estimator}

Previously, we compared the stopped process estimator and the naive estimator in terms of their systematic error and concluded that the stopped process estimator has a lower bias.  This raises a question: does this bias come at a trade-off of increased estimation error?  Indeed, given that the stopped process is a more complex estimator, one might expect there to be a bias-variance tradeoff between the two estimators.

However, analyzing the terms in the condition number suggests that this is not the case.
Consider two MSMs constructed using the same dataset, one using the naive estimator and the other using the stopped-process.
Both MSMs have the same sampling measure.  
However, the probability of leaving the domain is greater for the stopped-process estimator: whereas the naive estimator only checks if the trajectory has left at the lag time $\tlag$, the stopped process also checks for all times the trajectory leaves in the time interval $\tlag$.
This means the stopped process estimator will leave the domain earlier, causing~\eqref{eq:interpretation_term_2} to be smaller than for the naive estimator.
This is not a rigorous argument as we have ignored~\eqref{eq:interpretation_term_1}.  Moreover, the condition number is itself only a heuristic estimate of the noise-sensitivity of the MSM estimates\cite{thiede2015sharp}.  
However, it does suggest that the stopped process estimator is not necessarily more sensitive to noise than the naive estimator.

\section{Discussion}
\label{sec:discussion}
Our results highlight the importance of error reduction in constructing MSMs.  
We have divided the error into two sources: approximation error and estimation error.
Approximation error  is the error that is inherent to approximating the system's true dynamics using the dynamics of the Markov state model.
It  can be reduced by increasing the number of basis functions and by using the stopped process estimator, which reduces the bias that MSM estimates of the rate incur at large lag times.
Indeed, extensive work has been put into developing better MSM states\cite{li2016effect,husic2017ward} and using the stopped process estimator mitigates any additional error caused by improper choice of lag time.

However, this still leaves estimation error: the statistical error that occurs due to the finite number of samples used to estimate the averages in an MSM.
Reductions to the estimation error are limited by the amount of data available.  For instance, when constructing MSMs on trajectories from molecular dynamics simulations, collecting more data can take weeks or even months of compute time.  
Moreover, prior experiments comparing methods of estimating the committors and mean first-passage times have demonstrated that changing datasets can have a much stronger effect on the error than changing the estimator\cite{thiede2019galerkin}.  This motivates our search for heuristics that guage how sensitive a MSM is to statistical errors.

To this end, we have introduced the condition number, a well-known tool in numerical analysis, to Markov state models as a sensitivity metric.
While the condition number is a metric for sensitivity, it can still give valuable heuristics into choosing good MSMs.
For instance, analysis of the condition number highlights that sampling from the equilibrium measure can lead to large estimation errors.
Instead, using sampling measures that place more weight on important, high-energy states such as transition states can lead to more accurate MSMs.

\subsection{Implications for the variational principle for committors}

Our results on how the sampling measure affects estimation error also has interesting implications for recent work attempting to find the committor function using a variational principle.
Recent work has proposed estimating the committor by minimizing the loss functional\cite{krivov2013reaction,roux2021string,chen2023discovering}
\begin{equation}
    V[v] = \E_{\pi}[(v(X_\tlag) - v(X_0))^2]
    \label{eq:variational-functional}
\end{equation}
where $v$ is a function restricted to be 1 at the product state $B$ and 0 at the reactant state $A$.
By minimizing over possible $v$'s that obey the constraints, one can seek to find the committor function.
This approach can be applied to nonlinear parameterizations of the committor, such as neural networks.
Consequently, we expect that minimizing the variational principle can lead to small approximation errors: 
it allows us to use powerful representations that can represent the true committor to high accuracy.
Indeed, a similar approach has been highly successful for estimating slow modes of chemical systems\cite{noe2013variational,mardt2018vampnets,chen2019nonlinear}.

However, even if the approximation error is small the estimation error may still be large.
We suspect that this may be the case for committors constructed by minimizing~\eqref{eq:variational-functional} if the system has high potential energy barriers or transition states.
This is because the expectation must be taken over the equilibrium measure: if we replace $\pi$ in~\eqref{eq:variational-functional}
with a different probability measure, the functional is no longer minimized by the committor.
Our results from studying MSMs suggests that committor estimates constructed by sampling from equilibrium 
can lead to very large estimation errors,
since important states such as transition states and states on the reactive pathways are undersampled at equilibrium.

Indeed, our analysis does quantify the estimation error in a specific case.
In the supplementary information we show that if the committor is parameterized as a sum of indicator functions as in~\eqref{eq:committor_sum_of_indicators},
then minimizing~\eqref{eq:variational-functional} is mathematically equivalent to sampling initial points from the equilibrium measure, 
constructing an MSM with states $S_j$, and computing its committor. 
Under these conditions our results can be applied directly.
A detailed analysis of the role of estimation error when using this variational approach for arbitrary parameterizations of the committor is left for future work.
However, these initial results suggest if there are important intermediate states or other important states that are undersampled in the equilibrium measure, one should apply the variational principle with caution.

\subsection{Implications and future work}

While Markov State models are powerful tools for predicting rates and other dynamical quantities, empirical results suggest that they are limited by the amount of data available.  Our work has suggested simple heuristics for estimating how sensitive MSM estimates are to statistical errors: the condition number and a previously published 
bound on the asymptotic variance of the committor estimate.  While these bounds are loose, they nevertheless
give us insight into how statistical error in individual MSM entries can propagate to errors in dynamical quantities.
In particular, they highlight the importance of sampling intermediate states that connect reactant and product states.

Our results also leave possibilities for future work that both refines and uses our error analyses.
We have used condition numbers to measure how sensitive an estimated committor or mean first-passage time is to errors in the averages used to construct the MSM.  
However, by itself the condition number doesn't tell us about the magnitude of these errors.
Combining our work with prior efforts to model the distribution of errors in the entries of an MSM~\cite{singhal2005error,trendelkamp2015estimation} could give us a more complete picture of the error.
Moreover, our error analysis throughout this paper has focused on bounding the effect of any possible error on the MSM.
In practice, this strategy may be overly conservative: perturbations that lead to the largest errors may not be statistically representative of the errors that actually occur in practice.
By considering bad typical errors, rather than the worst-case errors, we may be able to develop probabilistic error bounds that more accurately reflect the true error in an MSM.
Lastly, while our primary intellectual contribution is in given intuition towards the causes of error in MSM estimates,
having mathematically motivated metrics for error sensitivity may open up the possibility for better algorithms that can adapt the sampling measure to reduce MSM error.


\begin{acknowledgement}
We would like to thank Lukas Stetzl, Charlotte Cheng, and Jonathan Weare for their helpful comments and support.  Code for evaluating the bound on the asymptotic variance of the committor was provided by Charlotte Cheng.
Jonathan Weare pointed out the importance of sampling measure in these error estimates, inspiring much of our analysis.
YT was funded by the Deutsche Forschungsgemeinschaft (DFG, German Research Foundation) - Project number 233630050 - TRR 146.  EHT and YT were funded by the Cornell University.  The Flatiron Institute is a division of the simons foundation.

\end{acknowledgement}

\clearpage
\begin{suppinfo}
\label{sec:supplement}
\subsection{Hyperparameters for Simulations}\label{ssec:hyperparams}
Here we present hyperparamateres for the numerical simulations provided in our paper.  In all simulations, the dynamicswere given by the overdamped Langevin dynamics, integrated using the overdamped BAOAB integrator from~\citet{leimkuhler2013rational} with a diffusion constant of unity and a timestep of $2.5 \times 10^{-4}$.

\subsubsection{The Triple Well}
Simulations were performed on a one-dimensional potential of form
\begin{equation*}
    U(x) = 1.5 x^6 - 3.2 x^4 + 2.74 x^2 - 0.12 x + 0.1.
\end{equation*}
MSMs were constructed using an even grid of 50 Markov states between -1.0 and 1.0, and this grid was re-used for both the stopped process and the naive estimator.

\subsubsection{The M\"uller-Brown Potential}
Simulations were performed on the M\"uller-Brown potential as presented in~\citet{muller1979location}, at a temperature of 30.
Markov States were constructed on an even grid of ten by ten Markov states, on a grid from $-1.25 < x < 1.0$ and $-.25 < y < 1.75$ respectively.  States that were not visited by any trajectory were discarded.  

\subsubsection{The Double Well potential}
Simulations were performed on a one-dimensional system with potential
\begin{equation*}
    U(x) =  x^4 -  x^2.
\end{equation*}
MSMs were constructed using 50 Markov states, chosen uniformly between the smallest and largest value of the trajectory observed in the simulation. The boundary states $\MSMB$ and $\MSMA$ consisted of the states closest to the first and second well, respectively.

\subsection{Stopped process derivation} 
Here, we present a simple derivation of the stopped process estimator in discrete time.
We begin by considering a generic linear system obeying the following operator equation with the stopped  transition operator; note that for a single step, the stopped transition operator and the true transition operator are the same.
\begin{equation}
(\SKoop{\tstep} - I)u = b
\end{equation}
If we multiply both sides by $(\SKoop{\tstep} - I)$, we have
\begin{equation}
(\SKoop{\tstep} - I)(\SKoop{\tstep} - I)u = (\SKoop{\tstep} - I)b 
\end{equation}
Rearranging terms and using the initial equation gives
\begin{equation}
    ((\SKoop{\tstep})^2 - I)u = (\SKoop{\tstep} + I)b
\end{equation}
Again multiplying both sides by $(\SKoop{\tstep} - I)$, we have
\begin{equation}
((\SKoop{\tstep})^3 - I)u = ((\SKoop{\tstep})^2 + \SKoop{\tstep}  + I)b \Delta t
\end{equation}
by the same manipulations.  
More generally, for multiply $n - 1$ times, we eventually arrive at the equations for the stopped process estimator.
\begin{equation}
((\SKoop{\tstep})^n - I)u = \left( \sum_{i=0}^{n-1} (\SKoop{\tstep})^i \right)b 
\end{equation}

\subsection{Connection between Markov State Models and the Variational Principle}

In the variational approach to finding the committor, the committor is defined as the function that minimizes the functional
\begin{equation}
\label{eq:variation-committor}
    q = \arg \min_{v \in S} \E_{\pi}[(v(X_\tau) - v(X_0))^2]
\end{equation}
where $S$ is a set of square-integrable functions that satisfy the boundary conditions
\begin{equation}
    v(x) = \begin{cases}
        0, & x \in A \\
        1, & x \in B.
    \end{cases}
\end{equation}
We first simplify 
\begin{equation}
    \begin{split}
\E_{\pi}[(v(X_\tau) -& v(X_0))^2] 
=
\int v(x)^2   \pi(x) \dd x  \\
&
- 2\int v(x) \Koop{\tstep} v(x) \pi(x) \dd x 
\\
&
+ \int \Koop{\tstep} v(x)^2 \pi(x) \dd x  
    \end{split}
\end{equation}
If the stationary distribution $\pi(x)$ is invariant under the transition operator $T$, the first and last terms are equal, and consequently
\begin{align}
\E_{\pi}&[(v(X_\tau) - v(X_0))^2] 
\\ &= 
2 \!\int\!\left(v(x)^2 - v(x) \Koop{\tstep} v (x)\right) \pi(x)\dd x  \\
&=
2 \!\int\! v(x) \left(\Iop - \Koop{\tstep} \right) v (x) \pi(x)\dd x 
\end{align}
and consequently,
\begin{equation}
    q(x) = \argmin_{v \in S}\!
\int\! v(x) \left(\Iop - \Koop{\tstep} \right) v (x) \pi(x)\dd x 
\end{equation}

Now, we restrict our attention to functions of the form 
$v = \sum\limits_{i} v_i \1_{S_i}(x)$.
Substituting and rearranging, the functional reduces to
\begin{equation}
    \sum_{i, j} v_i v_j \int \1_{S_i}(x) \left(\Iop - \Koop{\tstep} \right) \1_{S_j}(x) \pi(x)\dd x
\end{equation}
To find the extremum, we  differentiate with respect to the coefficients $v_n$
and set the result to zero:
\begin{align*}
    0 =& 
        \sum_i v_i \int \1_{S_i}(x) \left(\Iop - \Koop{\tstep} \right) \1_{S_n}(x) \pi(x)\dd x \nonumber \\ 
    &+ \sum_j v_j \int \1_{S_n}(x) \left(\Iop - \Koop{\tstep} \right) \1_{S_j}(x) \pi(x)\dd x 
\end{align*} 
If we further assume that the dynamics are reversible, then these two terms are equal, and our equation simplifies to
\begin{align*}
    0 =& \sum_j v_j \int \1_{S_n}(x) \left(\Iop - \Koop{\tstep} \right) \1_{S_j}(x) \pi(x) \dd x
\end{align*}
Finally, dividing both sides by $\int \1_{S_n}(x) \pi(x) \dd x$ and using the definition for the MSM transition matrix, we have that at the extremum the coefficients obey
\begin{equation}
    \sum_j T_{nj} v_j = v_n
\end{equation}
This is precisely the equation \eqref{eq:msm_committor} used to calculate the committor using an MSM.

\subsection{Derivations for the condition number}\label{ssec:cnumber_derivations}

Here, we present derivations for the probabilistic expressions presented in Subsection~\ref{ssec:condition_number_for_sensitivity}.
To derive~\eqref{eq:interpretation_term_1}, we observe that $\SCmat(\tlag)_{ij} = \mu_i \STmat_{ij}$.  Applying the definition of the induced matrix $1$-norm,
we have that
\begin{align*}
    \big\|  \SCmat(\tlag) & - \SCmat(0) \big\|_1 =
        \max_j \left[\sum_{i \in D}   \mu_i  \left|\STmat_{ij} - \delta_{ij} \right|
        \right]
        \\
&
    =\max_j \left[\sum_{i \in D, i \neq j}  \mu_i \STmat_{ij}  + \mu_j \left(1  - \STmat_{jj} \right) 
    \right] \\
&
    =\max_j \left[\sum_{i \in D, i \neq j}  \mu_i \STmat_{ij}  + \sum_{k \neq j} \mu_j  \STmat_{jk} 
    \right]
\end{align*}
The first term is $\P(\text{enter } j)$, the probability that in a single step of dynamics we observe the Markov chain enter $j$ from another state in $D$.  The second term is $\P(\text{leave }j)$, the probability that the dynamics starts in a single step of dynamics we see the chain start in $j$ and then go to any other state.

To derive~\eqref{eq:interpretation_term_2}, we again write the correlation matrix as a product of the sampling measure and the transition matrix,
\begin{align*}
    \left(\SCmat(\tlag) - \SCmat(0) \right)^{-1} =&
    \left(M (\STmat  - I)\right)^{-1}
    \\
    =&\left(\STmat  - I\right)^{-1} M^{-1},
\end{align*}
where $M$ is the matrices with $\mu$ on the diagonal for every state in $\MSMD$.
To simplify this, we employ the geometric series for matrices,
\begin{equation*}
    (\STmat - I )^{-1} = \sum_{l=0}^\infty (\STmat)^l.  
\end{equation*}
The $ij$'th entry of this matrix is, by definition, the probability that we see the system in state $j$ at time $lt$ after starting in $i$, summed over all states.
\begin{equation*}
    \sum_{l=0}^\infty (\STmat)^l =  \E \left[\sum_{s=0}^\infty \1_j (Y_{ls}) | Y_0= i  \right].
\end{equation*}
Consequently, we can rewrite the term as 
\begin{align*}
    &\left\|\left(\SCmat(\tlag) - \SCmat(0) \right)^{-1}\right\| \\
    &\qquad = 
    \max_j \frac{1}{\mu_j } \sum_i \E \left[\sum_{s=0}^\infty \1_j (Y_{ls}) | Y_0= i  \right].
\end{align*}

\begin{figure*}
    \centering
    \includegraphics[width=\linewidth]{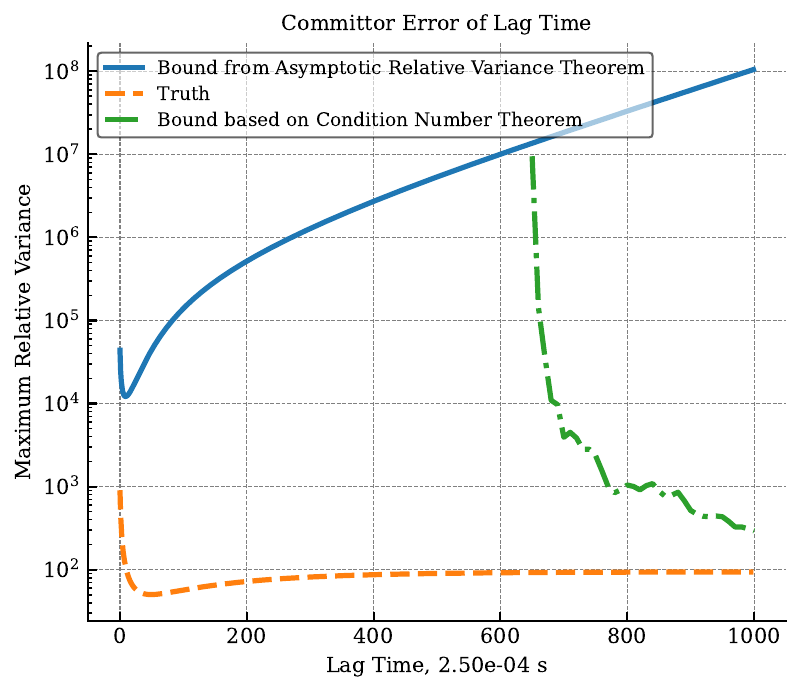}
    \caption{Comparison of the committor error bounds: the blue curve represents the estimation committor error bound based on the asymptotic variance presented in~\citet{cheng2024surprising}. The blue bound was estimated using the \textit{reference} transition matrix for the Müller-Brown system, whereas the green bound was calculated from both the \textit{estimated} transition matrix derived from a simulated trajectory and the \textit{reference} transition matrix for the same system.}

    \label{fig:compare-bounds}
\end{figure*}

\begin{table*}[p]
    \centering
    \begin{tabularx}{\textwidth}{|c|X|}
        \hline
        Symbol & Meaning \\
        \hline
        $X_t$ & The true dynamics of the system \\
        $Y_t$ & The dynamics of the MSM \\
        $\Koop{t}$ & Transition, or Koopman operator of the base process at lag time $\tlag$ \\
        $\SKoop{t}$ & Transition, or Koopman operator of the stopped process at lag time $\tlag$ \\
        $A$, $B$ & Sets in phase space, typically corresponding to reactants or products. \\
         $T^t$ &  True transition matrix from dynamics $X_t$  \\
         $\overline{T}^t$ &  Approximate transition matrix from time pairs $\{(X_0^n, X_t^n)\}_{n=1}^N$  \\
         $C^t$ & 
         True count matrix from dynamics $X_t$
         \\
         $\overline{C}^t$ & Approximate count matrix from time pairs $\{(X_0^n, X_t^n)\}_{n=1}^N$ 
         \\
        $\pi(x)$ & Stationary distribution of the Markov process \\
        $\mu(x)$ & Sampling distribution of the initial states of the MSM \\
        $q(x)$ & The forward committor function \\
        $m(x)$ & The mean-first passage time \\
        $\truestopt{A}$ & Stopping time for the Markov Process: the first time at which the system enters $A$, a set of points in the underlying state space. \\
        $\msmstopt{A}$ & Stopping time for the MSM Dynamics: first time at which the system enters $\MSMA$, $\MSMB$. \\
        $\MSMA$, $\MSMB$ & a collection of MSM states that cover $A$ or $B$, respectively. \\
        $\tstep$ & Time step of the underlying discrete-time Markov process \\
        \hline
    \end{tabularx}
    \caption{Symbols commonly used in this paper}
    \label{tab:common-symbols}
\end{table*}

\end{suppinfo}

\clearpage

\bibliography{msm-bibliography}
\end{document}